\documentclass[reprint,amsmath,amssymb,pra,aps,10pt,superscriptaddress]{revtex4-2}

\usepackage{graphicx}
\usepackage{dcolumn}
\usepackage{bm}
\usepackage{subcaption}
\usepackage{caption}
\usepackage{hyperref}
\usepackage{color}
\usepackage[dvipsnames]{xcolor}
\usepackage{placeins}
\usepackage[english]{babel}
\usepackage{amsmath}
\usepackage{indentfirst}
\usepackage{siunitx}
\usepackage{enumitem}
\usepackage[normalem]{ulem}   
\usepackage{ragged2e}     
\hypersetup{colorlinks,linkcolor={blue!70!black},citecolor={blue!70!black},urlcolor={black}}
\DeclareSIUnit{\gauss}{G}
\DeclareCaptionJustification{justified}{\justifying}
\renewcommand{\imath}[0]{i}
\newcommand{\Mtz}[1]{\left(\begin{matrix}#1\end{matrix}\right)}
\newcommand{\ParagSh}[1]{\vspace{.2cm}\noindent\textbf{\textit{#1.\--- }}}

\begin{document}

\preprint{APS/123-QED}

\title{Optical repumping and atom number balancing in a two-color MOT}

\author{Shubha~Deutschle}\email{shubha.deutschle@uni-tuebingen.de}
\affiliation{Physikalisches Institut, Eberhard Karls Universität Tübingen,
Auf der Morgenstelle 14, D-72076 Tübingen, Germany}

\author{L\H{o}rinc~S\'ark\'any}
\affiliation{Physikalisches Institut, Eberhard Karls Universität Tübingen,
Auf der Morgenstelle 14, D-72076 Tübingen, Germany}

\author{Mil\'an~J\'anos Negyedi}
\affiliation{Physikalisches Institut, Eberhard Karls Universität Tübingen,
Auf der Morgenstelle 14, D-72076 Tübingen, Germany}

\author{J\'ozsef~Fort\'agh}
\affiliation{Physikalisches Institut, Eberhard Karls Universität Tübingen,
Auf der Morgenstelle 14, D-72076 Tübingen, Germany}

\author{Andreas~G\"unther}
\affiliation{Physikalisches Institut, Eberhard Karls Universität Tübingen,
Auf der Morgenstelle 14, D-72076 Tübingen, Germany}

\author{Philippe~Wilhelm~Courteille}
\affiliation{Physikalisches Institut, Eberhard Karls Universität Tübingen,
Auf der Morgenstelle 14, D-72076 Tübingen, Germany}
\affiliation{Instituto de F\'isica de S\~ao Carlos, Universidade de S\~ao Paulo, S\~ao Carlos, SP 13566-970, Brazil}

\date{\today}

\begin{abstract}
\begin{description}
\item[Abstract] We study a novel repumping transition for $^{88}$Sr atoms trapped in a 'blue' magneto-optical trap. We show that, while the repumping efficiency is about three orders of magnitude smaller than for traditional schemes, it is sufficient for recycling all atoms, provided the repumping laser beams are arranged to form a 'green' magneto-optical trap (MOT) helping to cool and confine the atoms and preventing their loss. Our main findings are: (i) that the green MOT configuration is able to trap 10 times more atoms in the blue MOT than using the green transition merely as a repump, and (ii) that the atom numbers in the two-color MOT can be balanced through experimental control parameters. The interest of this scheme lies in its capability of reaching low temperature and its suitability for continuous atomic beam generation.
\end{description}
\end{abstract}

\maketitle

\section{\label{sec:Intro}Introduction}

Laser cooling and trapping plays a fundamental role in the growing field of quantum technologies, in particular, in precision metrology~\cite{Nemitz16,Campbell17,Norcia19,Ludlow15}, optical clocks~\cite{Takamoto05}, continuous superradiant lasers~\cite{Bennetts17, Dubey25}, quantum computers~\cite{Cao24}, (continuous) Bose-Einstein-condensation~\cite{Stellmer09,Stellmer10}, quantum simulations~\cite{Gross17, Bloch12, Bloch08, Dunning16}, quantum sensing~\cite{Degen17}, and molecular physics~\cite{McGuyer15}.

\begin{figure}[t]
    \includegraphics[width=\columnwidth]{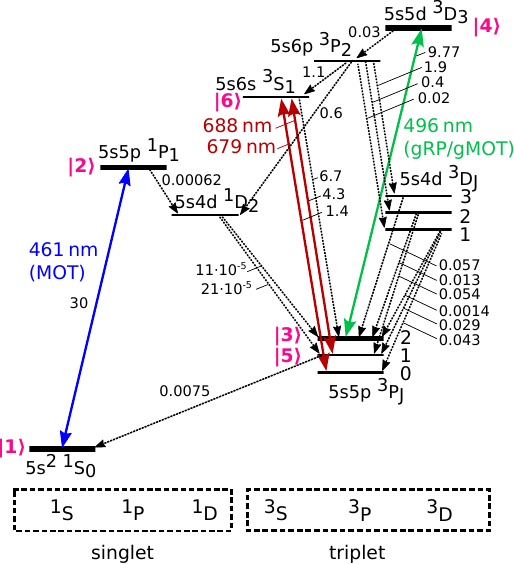}
    \caption{\small{Relevant energy levels of $^{88}\text{Sr}$. Colored arrows mark laser driven transitions, black arrows mark decay channels with their corresponding decay rates given in units of $2\pi\,\si{\mega\hertz}$~\cite{Akatsuka21}. The most relevant states for this study are labelled with $|1\rangle$ - $|6\rangle$.}}
    \label{fig:Levelscheme}
\end{figure}

Strontium, like other alkaline earth metals, provides a broad cooling transition operating between the levels $|1\rangle\equiv 5s^2\,^1S_0$ and $|2\rangle\equiv 5s5p\,^1P_1$, which we will call the 'blue' transition because of its wavelength of $\SI{461}{\nano\meter}$ (see Fig.~\ref{fig:Levelscheme}). Due to its large linewidth of $\Gamma_{12}=2\pi\times\SI{30}{\mega\Hz}$ this transition features a high capture velocity and fast laser cooling down to the (relatively high) Doppler temperature limit of about $\SI{1}{\milli\kelvin}$. Cold strontium experiments employ this transition as a first cooling stage in magneto-optical traps (MOT).

In order to reach lower temperatures, many experiments resort to the $\Gamma_{15}=2\pi\times\SI{7.5}{\kilo\Hz}$ narrow transition at $\SI{689}{\nano\meter}$ between the singlet ground state $|1\rangle$ and the triplet state $|5\rangle\equiv 5s5p\,^3P_1$ allowing for optical cooling to temperatures close to the recoil limit at about $\SI{300}{\nano\kelvin}$~\cite{Takamoto05,Rivero23}. As the capture velocity of this narrow transition is very low, most cooling protocols use temporal sequences of laser-cooling using this and the blue transition.

The performance of the blue MOT is limited by atom losses arising from the nonclosed nature of the blue cooling transition, as the excited state can decay into states outside the cooling cycle. With a branching ratio of 1:50000, the state $|2\rangle$ can decay into the metastable state $|5\rangle$~\cite{XuX03} at a rate of $\Gamma_{25}=2\pi\times\SI{159}{Hz}$ or into the state $|3\rangle\equiv 5s5p\,^3P_2$ at a rate of $\Gamma_{23}=2\pi\times\SI{90}{Hz}$ (see Appendix~\ref{sec:AppDecays}). Atoms having decayed into state $|5\rangle$ decay within $\SI{21}{\micro\second}$ back into the ground state~\cite{Stellmer14b,Drozdowski97} thus returning into the cooling cycle. On the other hand, atoms shelved in $|3\rangle$, due to it's long lifetime of about $\SI{500}{\second}$~\cite{Yasuda04} are withdrawn from the cooling dynamics and lost from the magneto-optical trap. For this reason, in order to reach high atom numbers in the blue MOT, it is essential to optically pump the atoms from that dark state back into the ground state.

Several repumping schemes have been proposed and tested in the past. A very early and well established scheme drives the $\SI{707}{\nano\meter}$ transition $5s5p\,^3P_2\rightarrow 5s6s\,^3S_1$. However, the excited state of that transition can also decay into the extremely long-lived $5s5p\,^3P_0$, so that a second repump at $\SI{679}{\nano\meter}$ driving the transition $5s5p\,^3P_0\rightarrow 5s6s\,^3S_1$~\cite{Dineen99, XuX03} is commonly used. Alternative pumping schemes have been demonstrated~\cite{Bowden19,Stellmer14,Poli05,Sorrentino06,Stellmer09,Stellmer14b,Poli06,Schkolnik2019, Poli06b,Moriya18, HuF19, Camargo-17, Mickelson09, Samland24,Mills17, Kurosu92, ZhangS20, Takamoto2020, Barker15} (see Appendix~\ref{sec:alternativeRepumps}).

In this paper, we report and study a novel repumping scheme for the blue MOT exploiting the transition from the metastable state $|3\rangle$ to the state $|4\rangle\equiv 5s5d\,^3D_3$ at $\SI{496}{\nano\meter}$ with a linewidth of $\Gamma_{34}=2\pi\times\SI{9.77}{\mega\Hz}$. At first glance, this choice may seem surprising, because the decay of the excited state into the state $|5\rangle$ is forbidden to first order by selection rules, rendering repumping rather inefficient as compared. The resulting decay rate is calculated in Appendix~\ref{sec:AppDecays} to be about $\Gamma_{45}=2\pi\times\SI{26.3}{\kilo\Hz}$, which is about three orders of magnitude smaller than the decay rates used, e.g., by the $\SI{707}{\nano\meter}$ repump. On the other hand, as long as the return rate $\Gamma_{14}$ is sufficiently large compared to the blue subsystem decay rate $\Gamma_{23}$, the pumping scheme is expected to operate. Moreover, manipulating $\Gamma_{14}$ by means of an additional repump laser at 688 nm coupling the levels $|5\rangle$ and $|6\rangle=5s6s\,^3S_1$, we demonstrate that the balance between the atomic populations evolving in the blue ($|1\rangle$,$|2\rangle$) and green ($|3\rangle$,$|4\rangle$) subsystems can be adjusted. Therefore the $\SI{688}{\nano\meter}$ laser represents a continuous control parameter for the equilibrium of atom numbers in the green and blue MOT.

Although the new repumping scheme is expected to be effective, additional mechanisms dominating the rate $\Gamma_{14}$ occur. Such a mechanism is given by atomic loss processes removing atoms from the trapping region. Indeed, during their absence from the blue MOT transition the atoms may drift away, if they are not trapped in the green subsystem. We show that these losses can efficiently be mitigated by a particular laser beam configuration realizing a second MOT on the 'green' transition at $\SI{496}{\nano\meter}$. In fact, there are good arguments for operating a green MOT on that transition: (i)~The green transition is relatively closed; (ii)~the transition linewidth is smaller than that of the blue MOT leading to lower Doppler temperatures; (iii)~the non-zero electronic spin of the ground and excited levels generate a Zeeman substructure allowing for polarization gradient cooling leading to sub-Doppler temperatures down to $\SI{45}{\micro\kelvin}$~\cite{Negyedi25}; (iv)~the green MOT does not share levels with the blue MOT, which decouples their dynamics to a large extend and enables their simultaneous operation~\cite{Akatsuka21,Negyedi25}; (v)~the optimum magnetic field gradients of the blue and green MOTs can be made to match such as to enable their simultaneous operation (see Appendix~\ref{sec:AppExternal}).

\bigskip

This paper is organized as follows: In Sec.~\ref{sec:Exp} we describe our two-color MOT experiment emphasizing the two different green laser 
configurations tested. After that, in Sec.~\ref{sec:Results}, we report on our measurements on atom numbers in the blue MOT with the green laser beams in MOT configuration (we will call this configuration gMOT) and in repump configuration (gRP). We also show how to control the balance between both MOTs via the detuning and intensity of the $\SI{688}{\nano\meter}$ repump laser. A simple theoretical model presented in Sec.~\ref{sec:Model} allows us to interpret an observed increase of the atom number in the blue MOT by one order of magnitude by just changing the beam configuration of the green laser such as to form (or not to form) a MOT on the green transition.
\begin{figure}[t]
    \includegraphics[width=8.7cm]{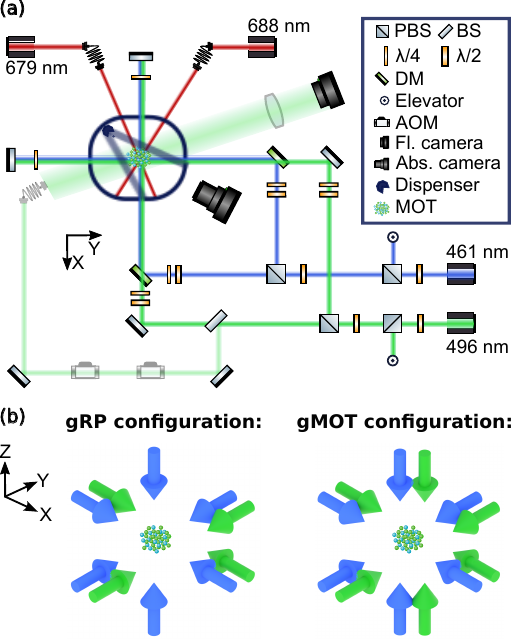} \caption{\small{Optical Setup. (a)~Schematics of the optical setup and detection units. PBS: polarizing beamsplitter, DM: dichroic mirror, BS: beam sampler, Fl./Abs. camera: Fluorescence/Absorption imaging camera. The elevators create the cooling beams in $z$-direction. All lasers are irradiated continuously, unless stated otherwise. (b)~Laser beam configuration for the gRP (left) and gMOT (right) configurations. In figures (a) and (b), gravity is directed along the $z$-direction.}}
    \label{fig:Setup}
\end{figure}

\section{\label{sec:Exp}Experiment}

\subsection{\label{sec:ExpSetup}Experimental Setup}

\ParagSh{Vacuum system and optical design} The experiment takes place in a small vacuum chamber. A pair of coils operated in anti-Helmholtz configuration is placed inside the vacuum chamber and generates the magnetic quadrupole field for the MOT with a field gradient $B^\prime$ along the symmetry axis (direction of gravity) of $\SI{30}{G/\centi\meter}$ per Ampère of current flowing through the coils. The blue MOT is directly and continuously loaded from a strontium dispenser without any precooling procedure.

The optical setup of the system is illustrated in Fig.~\ref{fig:Setup}. The blue ($\SI{461}{\nano\meter}$) and green ($\SI{496}{\nano\meter}$) beam is each divided into three beams using polarizing beam splitter cubes. The polarization of each single beam is controlled separately to achieve circular polarisation using combinations of quarter- and half-wave plates. The blue and green beams are superposed pairwise on dichroic mirrors. The three beams pass through the chamber orthogonally. Behind the chamber each beam passes an achromatic quarter-wave plate and is retroreflected by a mirror. The atomic cloud is detected by either imaging its fluorescence on a CMOS camera (UI-1240SE-NIR-GL from IDS with a lens system) or by an absorption imaging using an achromatic doublet in a $2f$-$2f$ configuration. To measure the green and blue fluorescence separately, we place short, respectively, long pass filters in front of the camera. 

The powers and beam waists used in the experiment are listed in Appendix~\ref{sec:AppSatint}. The blue cooling laser is tuned below resonance, $\Delta_{12}=-2\Gamma_{12}$. All lasers used for this work are frequency-stabilized to a wavemeter (HighFinesse WS8-2). Further details of the experimental setup can be found in~\cite{Negyedi25}.

\ParagSh{Green beam setup} If operated as a single incident beam, radiation pressure exerted by a highly cycling repump would push the atoms out of the blue MOT region more than other commonly used schemes. To mitigate this effect, we design our repump beam path in one of two different retroreflected beam configurations that we call gRP and gMOT, see Fig.~\ref{fig:Setup}(b). In gRP configuration, four beams are adjusted to pairwise counterpropagate and cross the blue MOT region orthogonally. This eliminates active acceleration of the atoms in the horizontal x-y plane. In gMOT configuration six green laser beams are aligned to pairwise counterpropagate and cross the blue MOT. For both configurations, in total the same green and blue laser powers are irradiated on the atoms.
\begin{figure*}[t]
    \centering
    \includegraphics[width=0.85\textwidth]{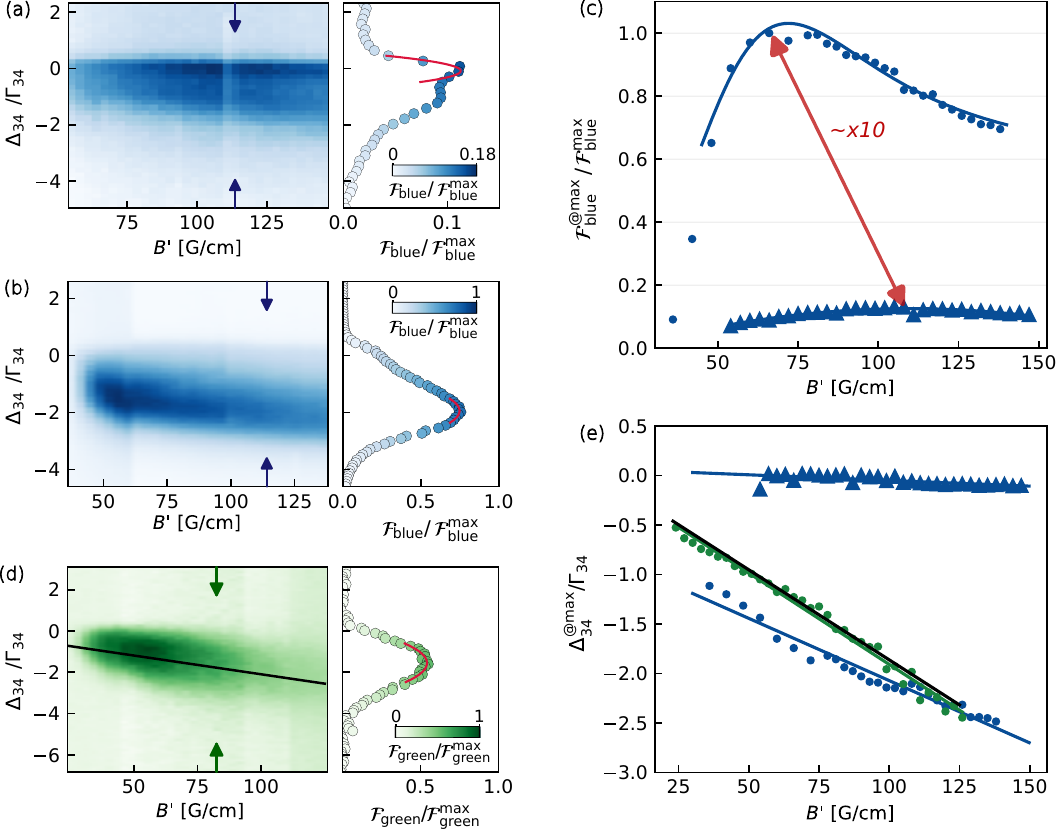}
    \caption{\small{Fluorescence measurements. Blue fluorescence $\mathcal{F}_\text{blue}$ measured at $\SI{461}{\nm}$ with (a)~the green lasers in gRP configuration and (b)~in gMOT configuration: (left panels)~Blue fluorescence normalized to its maximum value in gMOT configuration as a function of the magnetic field gradient and green laser detuning; (right panels)~Exemplary linescans (blue circles) for a magnetic field gradient of $B'=\SI{114}{G/cm}$ [marked by arrows in the left panels]. The red solid lines show parabolic fits used for extracting the field-gradient-depending detuning at which the fluorescence reaches it's maximal value $\mathcal{F}_\text{blue}^\text{max}(B')$. (c)~Maximal blue fluorescences $\mathcal{F}_\text{blue}^\text{@max}$ normalized to its maximum value in gMOT configuration $\mathcal{F}_\text{blue}^\text{max}$ as a function of the magnetic field gradient with the green lasers in gMOT configuration (dots) or in gRP configuration (triangles). The solid lines are polynomial fits to guide the eye. (d)~Green fluorescence $\mathcal{F}_\text{green}$ measured at $\SI{496}{\nm}$. The black solid line indicates the green laser detuning at which the green MOT potential is deepest as a function of the magnetic field gradient [see Eq.~(\ref{eq:App23})]. (right)~Exemplary linescan (green circles) for a magnetic field gradient of $B'=\SI{82}{G/cm}$ [marked by arrows in the left panel]. (e)~Field-gradient-depending green laser detunings $\Delta_{34}^{\text{@max}}$ at which the fluorescences are maximal: (blue dots) blue fluorescence with the green laser in gMOT configuration, (blue triangles) blue fluorescence with the green laser in gRP configuration, (green dots) green fluorescence with the green laser in gMOT configuration and additionally the $\SI{688}{\nano\meter}$ laser, (black) theoretical curve as calculated from Eq.~(\ref{eq:App23}). The blue solid lines are linear fits to guide the eye. Note, that all spectra have been shifted by $\Delta_{34}/\Gamma_{34}=0.28$ to correct for systematic errors.}}
    \label{fig:RP_measurement}
\end{figure*}

\subsection{\label{sec:TwoColorMOT}Interplay between blue and green subsystems}

The state $|5\rangle$ plays a pivotal role in balancing between the blue and green subsystems, as all green-to-blue decay paths $|4\rangle\rightarrow|1\rangle$ and one of the blue-to-green decay paths $|2\rangle\rightarrow|3\rangle$ transit through this state. Manipulating it's population one can control the atom number ratio of the two subsystems. This can be achieved by an additional laser at $\SI{688}{\nano\meter}$, which pumps the atoms via the excited state $|6\rangle$ into the metastable state $|3\rangle$ belonging to the green subsystem. The depletion of the state $|5\rangle$ inhibits the green-to-blue decay channel thus closing the green subsystem. In contrast, the reverse blue-to-green decay comprises one path, $|2\rangle\rightarrow 5s4d\,^1D_2\rightarrow |3\rangle$, which bypasses the state $|5\rangle$, such that the blue-to-green decay cannot be set to zero. Since it is the ratio of blue-to-green and green-to-blue decay which rules the equilibrium between the two subsystems, it is possible to control the equilibrium by tuning the intensity of the $\SI{688}{\nano\meter}$ pump.

\bigskip

So far we only considered internal pump processes, but transitions from green-to-blue can also be driven via external dynamics: During their time in the green subsystem with the green laser set to gRP configuration, the atoms move outside the spatial region where the blue MOT provides cooling and trapping forces. Thus, atoms staying too long in the green subsystem get lost. The situation is different when the green subsystem is set to gMOT configuration, as this confines the atoms and closes the external loss channel.

\section{\label{sec:Results}Measurements}

\subsection{\label{sec:Blue MOT}Fluorescence of the blue and green MOT}

\ParagSh{Blue MOT fluorescence} To operate the blue MOT we irradiate all lasers shown in Fig.~\ref{fig:Setup}(a) except for the $\SI{688}{\nano\meter}$ beam. The equilibrium between the blue and green subsystems is then located on the blue side. The green laser is aligned either in gRP or in gMOT configuration [see Fig.~\ref{fig:Setup}(b)]. To compare the repumping efficiency of both configurations, we sweep the green laser frequency detuning $\Delta_{34}$ for different magnetic field gradients $B'$ while measuring the blue MOT fluorescence $\mathcal{F}_\text{blue}$. The measurement results are shown in Figs.~\ref{fig:RP_measurement}(a,b). In this setup (without the $\SI{688}{\nano \meter}$ laser) we are not able to detect any green fluorescence. Exemplary linescans at a magnetic field gradient of $B'=\SI{114}{G/cm}$ are shown at the right sides of panels (a) and (b). Both data sets are normalized to the maximal fluorescence $\mathcal{F}_\text{blue}^\text{max}$ of the data set obtained in gMOT configuration.

\bigskip

Further, we analyze the data of Figs.~\ref{fig:RP_measurement}(a,b) by extracting the maximal blue fluorescence for every magnetic field gradient yielding to Fig.~\ref{fig:RP_measurement}(c). Comparing them, we find that in gMOT configuration the blue fluorescence is almost up to 10 times larger than in gRP configuration. Since the blue MOT fluorescence is proportional to the atom number in the blue MOT, we infer that 10 times more atoms are trapped in the blue MOT when the green beams are in gMOT configuration. We attribute this to a higher repumping efficiency of the green laser in gMOT configuration due to a closure of the external loss channel. 

\ParagSh{Green MOT fluorescence} To make green fluorescence visible, all lasers shown in Fig.~\ref{fig:Setup}(a) including the $\SI{688}{\nano\meter}$ laser are in operation. The green laser is adjusted in gMOT configuration. The fluorescence of the green MOT $\mathcal{F}_\text{green}$ is measured while the green laser is varied in frequency for several magnetic field gradients. The data are shown in Fig.~\ref{fig:RP_measurement}(d), an exemplary linescan at a magnetic field gradient of $B'=\SI{82}{G/cm}$ is shown at the right side.
\begin{figure*}[htbp]
    \centering
    \includegraphics[width=\textwidth]{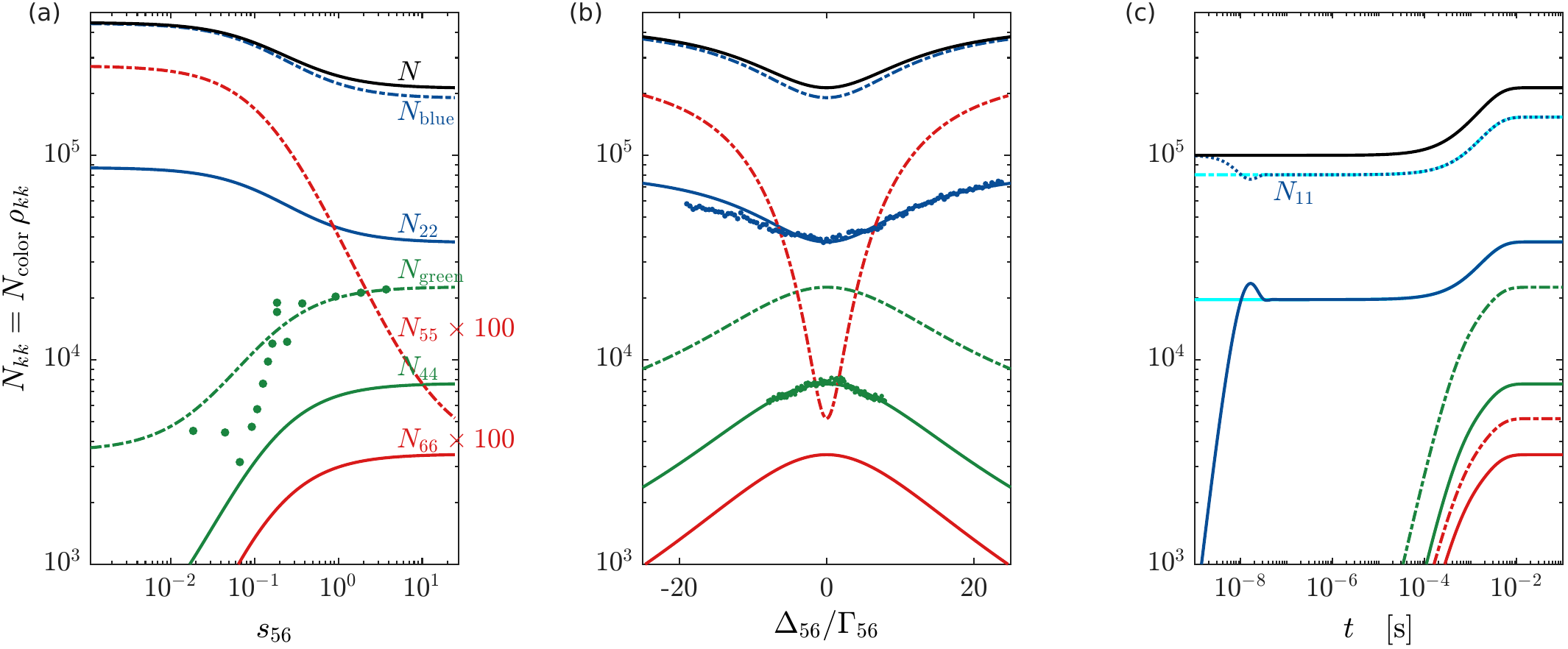}
    \caption{\small{Simulated steady state atom numbers $N_{kk}$ in the states $|2\rangle$, $|4\rangle$, $|5\rangle$, and $|6\rangle$, populations of the two subsystems $N_\text{blue}=\sum_{k=1}^2N_{kk}$, $N_\text{gr:rd}=\sum_{k=3}^6N_{kk}$, and total atom number of the system $N=N_\text{blue}+N_\text{gr:rd}$. (a)~Green dots denote atom numbers in the green MOT determined by absorption measurement $N_\text{gr:rd}$ as a function of the saturation of the $\SI{688}{\nano\meter}$ laser. All solid and dash-dotted lines are simulated with the full Bloch equations [Eq.~(\ref{eq:mod03})]. The hybrid Bloch-rate equations model yield identical results.
    (b)~ Frequency scan of the $\SI{688}{nm}$ laser showing the same level populations as in panel (a). The blue/green dots are atom numbers as derived from blue/green fluorescence measurements.
    (c)~Time evolution of the populations in the blue and green subsystems as denoted by their respective colors. Same color and linestyle coding as in panels (a,b). The cyan lines are obtained from the hybrid Bloch-rate equations model starting at $t=0$ in equilibrium of the blue subsystem. The initial atom number assumed in the ground state is set to $N_{11}=10^6$, to generate visible Rabi oscillations in the blue subsystem helping to illustrate separation of time scales for intra-subsystem equilibrium and inter-subsystem pumping. For all plots, the detunings have been set to match the experimental conditions and are (unless varied) $\Delta_{12}=-\Gamma_{12}/2$, $\Delta_{34}=0$, $\Delta_{56}=0$. The remaining parameters have been adjusted to match the experimental data: $s_{12}=1.3$, $s_{34}=2.1$, $s_{56}=25$,  $R_\text{load}=10^8\,\text{s}^{-1}$, $\Gamma_\text{blue}=190\,\text{s}^{-1}$, and $\Gamma_\text{gr:rd}=2500\,\text{s}^{-1}$. The saturation parameters agree with measurements of power and beam waist.}}
    \label{fig:Balancing}
\end{figure*}

\ParagSh{Discussion} Fig.~\ref{fig:RP_measurement}(e) shows the green laser detunings $\Delta_{34}/\Gamma_{34}$ at which the maximal fluorescence was observed. The blue dots and triangles refer to the blue fluorescence in gMOT and gRP configuration of panels (a,b), respectively. The green dots refer to the green laser detuning for maximal green fluorescence. The solid lines show linear fits to the data revealing slopes of $-120\,\text{kHz}/\text{G/cm}$ for the blue fluorescence in gMOT configuration, $-11\,\text{kHz}/\text{G/cm}$ for gRP configuration, and $-180\,\text{kHz}/\text{G/cm}$ for the green fluorescence in gMOT configuration. The fact that the slope is close to zero in gRP configuration is understood as absence of any three-dimensional MOT dynamics (see Appendix~\ref{sec:AppExternal}). On the other hand, in gMOT configuration the slope is similar to the one for the green fluorescence indicating the existence of a green MOT, although it is not detectable in our experiment.

\subsection{\label{sec:Green MOT}Controlling the blue-green MOT balance}

The balancing between the blue and green subsystem in the two-color MOT can be controlled by the intensity and detuning of the $\SI{688}{\nano\meter}$ light with respect to the $|5\rangle-|6\rangle$ transition. For demonstration, we have measured the atom number in the green MOT $N_\text{green}$ via absorption imaging as a function of the saturation parameter of the $\SI{688}{\nano\meter}$ beam $s_{56}$. Results are presented in Fig.~\ref{fig:Balancing}(a). We also measure the fluorescence of the blue and green MOTs while sweeping the detuning $\Delta_{56}$ of the $\SI{688}{\nano\meter}$ laser and show the inferred atom number in Fig.~\ref{fig:Balancing}(b). To this end, the fluorescence is calibrated to atom number via absorption imaging. We observe that the atom number in the green MOT increases while the atom number in the blue MOT decreases and vice versa. The blue MOT cannot be completely emptied because the green MOT is being loaded through spontaneous decay from the state $|2\rangle$. Therefore the blue MOT must exist to load the green MOT. The measurements are described and understood in terms of a theoretical model outlined in Sec.~\ref{sec:Model}.

\section{\label{sec:Model}Theoretical modelling}

The atom numbers in both subsystems are proportional to the measured fluorescences $\mathcal{F}_\text{color}$ ($\text{color}\in\{\text{blue, green}\}$),
\begin{align}\label{eq:F_color}
    \mathcal{F}_\text{blue} & = \alpha_\text{blue} \,\Gamma_{12}\,N_{22}\\
    \mathcal{F}_\text{green} & = \alpha_\text{green}\,\Gamma_{34}\,N_{44}~.\nonumber
\end{align}
$\Gamma_{12}$ and $\Gamma_{34}$ are the linewidths of the blue and green MOT transitions, respectively, and $N_{22}$ and $N_{44}$ the occupations of the respective excited states. The proportionality factors $\alpha_\text{color}$ between observed fluorescence and atom number in the MOTs describe the collection solid angle and the detector efficiency. Their experimental calibration from independent measurements of the atom numbers $N_\text{color}$ (e.g.~via standard absorption imaging) based on Eq.~(\ref{eq:F_color}) would require knowledge of the populations $N_{22}$ respectively $N_{44}$ which, a priori, are unknown and must be estimated from a theoretical model.

\bigskip

\begin{figure}[htbp]
    \centering
    \includegraphics[width=8.7cm]{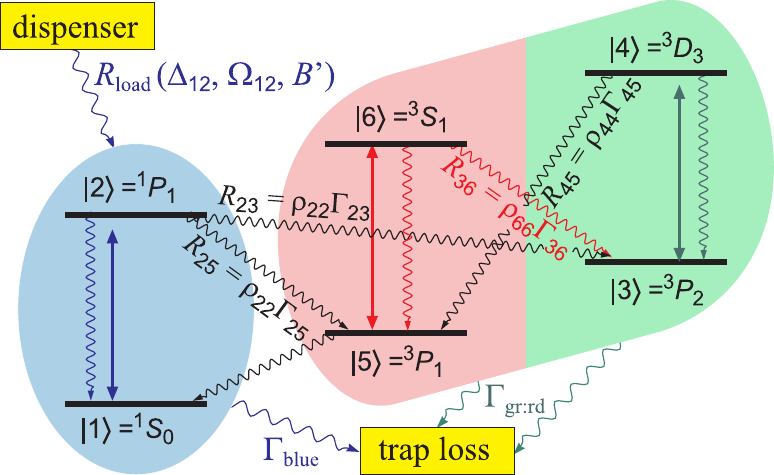} 
    \caption{\small{Illustration of the basic idea of the modelling. The atomic level scheme is described as three incoherently coupled two-level systems representing the blue ($|1\rangle-|2\rangle$), the green ($|3\rangle-|4\rangle$) and the red ($|5\rangle-|6\rangle$) transitions~\cite{Note1}. Only the blue subsystem can be adiabatically separated.}}
    \label{fig:TwoTwoTwoLevelScheme}
\end{figure}

For our complex multi-level system we propose a simple model taking account of the fact that the blue, green and red transitions are only incoherently coupled via spontaneous decay, as illustrated in Fig.~\ref{fig:TwoTwoTwoLevelScheme}. Here, for simplicity we ignore the $\SI{679}{nm}$ repump laser, which merely avoids population of the $5s5p\,^3P_0$ level. We begin setting up the six-level open Bloch equations focusing on the internal dynamics, but considering loading and loss processes from, respectively, into a reservoir. In a second step, we solve the Bloch equations separating processes occurring on very different time scales. Finally, we propose a simplistic model for the dependence of the MOT loading and loss rates on the detuning of the green laser $\Delta_{34}$ and on the magnetic field gradient $B'$.

\subsection{\label{sec:ModelDecoupling}Bloch equations for three incoherently coupled two-level systems}

Using an open system description that accounts for the system's loading and loss rates, we normalize the level populations $N_{kk}(t)$ and coherences $N_{k\ne l}$ to the instantaneous atom number $N(t)$ in the whole system, $\sum_kN_{kk}(t)=N(t)$. We choose a particular arrangement for the Bloch vector $\vec N$ and define an inhomogeneity vector $\vec b$ describing the loading of the double MOT system at a rate $R_\text{load}$ from a background reservoir,

\begin{equation}\label{eq:mod01}
\begin{scriptsize}
    \vec N = \Mtz{N_{11} \\ N_{22} \\ N_{12} \\ N_{21} \\ N_{55} \\ N_{66} \\ N_{56} \\ N_{65} \\ N_{33} \\ N_{44} \\ N_{34} \\ N_{43}} ~~~~~\text{and}~~~~~ 
    \vec b = \Mtz{R_\text{load} \\ 0 \\ 0 \\ 0 \\ 0 \\ 0 \\ 0 \\ 0 \\ 0 \\ 0 \\ 0 \\ 0}
    \end{scriptsize}
\end{equation}
and write the open Bloch equations for the three incoherently coupled two-level subsystems called the blue, red, and green subsystems as,
\begin{equation}\label{eq:mod02}
    \frac{d}{dt}\vec N = B\vec N+\vec b~.
\end{equation}
The stationary and time-dependent solutions are,
\begin{align}\label{eq:mod03}
    \vec N(\infty) & = -B^{-1}\vec b\\
    \vec N(t) & = e^{Bt}\vec N(0)+(\mathbb{I}-e^{Bt})\vec N(\infty)~.
\end{align}
For the Bloch and inhomogeneity vectors defined in Eq.~(\ref{eq:mod01}), the Liouvillean takes the form (see Appendix~\ref{sec:AppDerive}),\begin{widetext}\begin{scriptsize}\begin{align}\label{eq:mod04}
    & B =\\
    & \left(\begin{array}[c]{cccc|cccc|cccc}
    -\Gamma_\text{blue} & \Gamma_{12} & \tfrac{\imath}{2}\Omega_{12} & -\tfrac{\imath}{2}\Omega_{12} & \Gamma_{15} & 0 & 0 & 0 & 0 & 0 & 0 & 0\\
    0 & -\Gamma_{12}-\Gamma_{23}-\Gamma_{25} & -\tfrac{\imath}{2}\Omega_{12} & \tfrac{\imath}{2}\Omega_{12} & 0 & 0 & 0 & 0 & 0 & 0 & 0 & 0\\
    \tfrac{\imath}{2}\Omega_{12} & -\tfrac{\imath}{2}\Omega_{12} & -\Lambda_{12} & 0 & 0 & 0 & 0 & 0 & 0 & 0 & 0 & 0\\
    -\tfrac{\imath}{2}\Omega_{12} & \tfrac{\imath}{2}\Omega_{12} & 0 & -\Lambda_{12}^* & 0 & 0 & 0 & 0 & 0 & 0 & 0 & 0\\\hline
    0 & \Gamma_{25} & 0 & 0 & -\Gamma_{15}-\Gamma_\text{gr:rd} & \Gamma_{56} & \tfrac{\imath}{2}\Omega_{56} & -\tfrac{\imath}{2}\Omega_{56} & 0 & \Gamma_{45} & 0 & 0\\
    0 & 0 & 0 & 0 & 0 & -\Gamma_{56}-\Gamma_{36} & -\tfrac{\imath}{2}\Omega_{56} & \tfrac{\imath}{2}\Omega_{56} & 0 & 0 & 0 & 0\\
    0 & 0 & 0 & 0 & \tfrac{\imath}{2}\Omega_{56} & -\tfrac{\imath}{2}\Omega_{56} & -\Lambda_{56} & 0 & 0 & 0 & 0 & 0\\
    0 & 0 & 0 & 0 & -\tfrac{\imath}{2}\Omega_{56} & \tfrac{\imath}{2}\Omega_{56} & 0 & -\Lambda_{56}^* & 0 & 0 & 0 & 0\\\hline
    0 & \Gamma_{23} & 0 & 0 & 0 & \Gamma_{36} & 0 & 0 & -\Gamma_\text{gr:rd} & \Gamma_{34} & \tfrac{\imath}{2}\Omega_{34} & -\tfrac{\imath}{2}\Omega_{34}\\
    0 & 0 & 0 & 0 & 0 & 0 & 0 & 0 & 0 & -\Gamma_{34}-\Gamma_{45} & -\tfrac{\imath}{2}\Omega_{34} & \tfrac{\imath}{2}\Omega_{34}\\
    0 & 0 & 0 & 0 & 0 & 0 & 0 & 0 & \tfrac{\imath}{2}\Omega_{34} & -\tfrac{\imath}{2}\Omega_{34} & -\Lambda_{34} & 0\\
    0 & 0 & 0 & 0 & 0 & 0 & 0 & 0 & -\tfrac{\imath}{2}\Omega_{34} & \tfrac{\imath}{2}\Omega_{34} & 0 & -\Lambda_{34}^*\end{array}\right)\nonumber
\end{align}\end{scriptsize}\end{widetext}
with the abbreviations,
\begin{align}\label{eq:mod05}
    \Lambda_{12} & = \imath\Delta_{12}+\tfrac{1}{2}(\Gamma_{12}+\Gamma_{25}+\Gamma_{23})\\
    \Lambda_{34} & = \imath\Delta_{34}+\tfrac{1}{2}(\Gamma_{34}+\Gamma_{45})\nonumber\\
    \Lambda_{56} & = \imath\Delta_{56}+\tfrac{1}{2}(\Gamma_{56}+\Gamma_{36}+\Gamma_{15})\nonumber ~.
\end{align}
$\Gamma_\text{blue}$ and $\Gamma_\text{gr:rd}$ are loss rates from the blue and green-red subsystems, respectively. In the matrix Eq.~\eqref{eq:mod04} they are neglected wherever dominated by strong internal decay rates.
\bigskip

\ParagSh{Separation of time scales} Inspecting the Liouvillean $B$ we observe, that it almost breaks down into three two-level subspaces. While the blue and green-red subspaces are only coupled by small decay rates $\Gamma_{23},\Gamma_{25},\Gamma_{15}\ll\Gamma_{56}$, the green and red subspaces are coupled by very strong decay rates $\Gamma_{36}\approx \Gamma_{56}$. This means that the blue subsystem can be decoupled from the others, but the red and green subsystem cannot be decoupled from each other. This allows for the simplification presented in the next section.

The separation of time scales is visualized in Fig.~\ref{fig:Balancing}(c), which shows the time evolution of the populations of all three two-level systems as derived from Eq.~(\ref{eq:mod04}). The internal dynamics of atoms in the blue transition (blue solid and dash-dotted lines) appears as damped Rabi oscillations occurring on a time scale of about $\SI{10}{\nano\s}$, while pumping towards the green subsystem takes place between $\SI{0.1}{ms}$ and $\SI{10}{ms}$. With the parameters chosen for this simulation, the MOT loading and loss dynamics dominates the long term behavior after $\SI{1}{s}$.

\subsection{\label{sec:ModelSolution}Hybrid Bloch-rate equations}

The blue subsystem, consisting of levels $|1\rangle$ and $|2\rangle$ and characterized by the laser detuning $\Delta_{12}$ and the Rabi frequency $\Omega_{12}$, represents an almost closed system on the time scale of its own dynamics, which is governed by $\Gamma_{12}$. If it rapidly reaches equilibrium, we find for the normalized population in the blue subsystem,
\begin{equation}\label{eq:mod11}
	\rho_{22}(\infty) = \frac{1}{2\left(1+\frac{1}{s_{12}}\left(1+\left(\frac{2\Delta_{12}}{\Gamma_{12}}\right)^2\right)\right)}~.
\end{equation}

For the green-red subsystem we still need to solve a set of open Bloch equations with the reduced Liouvillean,
\begin{widetext}\begin{small}\begin{equation}\label{eq:mod12}
	M = \left(\begin{array}{cccc|cccc}
	    -\Gamma_{15} & \Gamma_{56} & \tfrac{\imath}{2}\Omega_{56} & -\tfrac{\imath}{2}\Omega_{56} & 0 & \Gamma_{45} & 0 & 0\\
        0 & -\Gamma_{56}-\Gamma_{36} & -\tfrac{\imath}{2}\Omega_{56} & \tfrac{\imath}{2}\Omega_{56} & 0 & 0 & 0 & 0\\
        \tfrac{\imath}{2}\Omega_{56} & -\tfrac{\imath}{2}\Omega_{56} & -\Lambda_{56} & 0 & 0 & 0 & 0 & 0\\
        -\tfrac{\imath}{2}\Omega_{56} & \tfrac{\imath}{2}\Omega_{56} & 0 & -\Lambda_{56}^* & 0 & 0 & 0 & 0\\\hline
        0 & \Gamma_{36} & 0 & 0 & 0 & \Gamma_{34} & \tfrac{\imath}{2}\Omega_{34} & -\tfrac{\imath}{2}\Omega_{34}\\
        0 & 0 & 0 & 0 & 0 & -\Gamma_{34}-\Gamma_{45} & -\tfrac{\imath}{2}\Omega_{34} & \tfrac{\imath}{2}\Omega_{34}\\
        0 & 0 & 0 & 0 & \tfrac{\imath}{2}\Omega_{34} & -\tfrac{\imath}{2}\Omega_{34} & -\Lambda_{34} & 0\\
        0 & 0 & 0 & 0 & -\tfrac{\imath}{2}\Omega_{34} & \tfrac{\imath}{2}\Omega_{34} & 0 & -\Lambda_{34}^*
    \end{array}\right) ~~~~~\text{with}~~~~~ 
    \vec\rho = \Mtz{\rho_{55} \\ \rho_{66} \\ \rho_{56} \\ \rho_{65} \\ \rho_{33} \\ \rho_{44} \\ \rho_{34} \\ \rho_{43}}~.
\end{equation}\end{small}\end{widetext}
The Bloch vectors for both, the blue and the green-red subsystems, are separately normalized to 1, that is, $\rho_{11}+\rho_{22}=1=\sum_{k=3}^6\rho_{kk}$.

\bigskip

Population transfer due to decay between the blue and the green-red (index gr:rd) subsystems, $R_{23}=\Gamma_{23}\rho_{22}$, $R_{25}=\Gamma_{25}\rho_{22}$, and $R_{15}=\Gamma_{15}\rho_{55}$, is now taken care of through rate equations,
\begin{equation}\label{eq:mod13}
    \frac{d}{dt}\Mtz{N_\text{blue} \\ N_\text{gr:rd}} = A\Mtz{N_\text{blue} \\ N_\text{gr:rd}}+\vec a
\end{equation}
where
\begin{equation}\label{eq:mod14}
    A = \Mtz{-\Gamma_\text{blue}-R_{23}-R_{25} & R_{15} \\ R_{23} & -\Gamma_\text{gr:rd}-R_{15}}
\end{equation}
and
\begin{equation}\label{eq:mod15}
    \vec a = \Mtz{R_\text{load} \\ 0} \, .
\end{equation}
\ParagSh{Steady state solution}
The steady state solution is,
\begin{small}\begin{align}\label{eq:mod16}
    & \Mtz{N_\text{blue}(\infty) \\ N_\text{gr:rd}(\infty)} = -A^{-1}\vec a\\
    & = \frac{R_\text{load}\Mtz{\Gamma_\text{gr:rd}+R_{15} \\ R_{23}}}{\Gamma_\text{blue}(\Gamma_\text{gr:rd}+R_{15})+\Gamma_\text{gr:rd}(R_{23}+R_{25})+R_{25}R_{15}}\nonumber
\end{align}\end{small}
and the total atom number,
\begin{align}\label{eq:mod17}
    N(t) & = N_\text{blue}(t)+N_\text{gr:rd}(t)\\
    N_{kl}(t) & = N_\text{blue}(t)\rho_{kl}(t) ~~~~~\text{for}~~ k,l=1,2\nonumber\\
    N_{kl}(t) & = N_\text{gr:rd}(t)\rho_{kl}(t) ~~~~~\text{for}~~ k,l=3,4,5,6\nonumber \, .
\end{align}

\ParagSh{Time-dependence of optical pumping} The rate equations ruling the optical pumping between the two subsystems can be solved analytically,
\begin{equation}\label{eq:mod18}
    \Mtz{N_\text{blue}(t) \\ N_\text{gr:rd}(t)} = e^{At}\Mtz{N_\text{blue}(0) \\ N_\text{gr:rd}(0)}+(\mathbb{I}-e^{At})\Mtz{N_\text{blue}(\infty) \\ N_\text{gr:rd}(\infty)}
\end{equation}
via diagonalization of the matrix $A$ given in Eq.~\eqref{eq:mod14}. An example for the time evolution of the populations obtained from Eq.~\eqref{eq:mod03} and Eq.~\eqref{eq:mod18} is shown in Fig.~\ref{fig:Balancing}(c). Both approaches yield identical results in the equilibrium.

\ParagSh{Balance between blue and green MOT} From Eq.~\eqref{eq:mod16} we estimate the steady-state balance between the blue and green-red subsystem, \begin{equation}\label{eq:mod19}
    \frac{N_\text{gr:rd}(\infty)}{N_\text{blue}(\infty)} 
    = \frac{\Gamma_{23}\rho_{22}(\infty)}{\Gamma_\text{gr:rd}+\Gamma_{15}\rho_{55}(\infty)} \, .
\end{equation}
The population $\rho _{22}(\infty )$ is determined by Eq.~(\ref{eq:mod11}). Inserting the decay rates $\Gamma_{15}$ and $\Gamma_{23}$, we thus estimate using $\rho_{22}(\infty) \leq 1/2$,
\begin{equation}\label{eq:mod20}
    \frac{N_\text{gr:rd}(\infty)}{N_\text{blue}(\infty)} \leq \frac{1}{2\Gamma_\text{gr:rd}/\Gamma_{23}+2 \Gamma_\text{15}/\Gamma_{23} \rho_{55}(\infty)}~.
\end{equation}
This means, that the balance between the blue and the green-red subsystems crucially depends on the population $\rho_{55}$. The atom number in the green-red system can thus be maximized by minimizing $\rho_{55}$, e.g.~via the intensity of a repumping laser. However, it will never exceed the value $N_\text{blue}(\infty)\Gamma_{23}/(2\Gamma_\text{gr:rd})$, which demonstrates the necessity of not loosing atoms in the green subsystem (e.g.~by trapping them in a green MOT).

\ParagSh{Discussion} The data in Fig.~\ref{fig:Balancing}(a,b) illustrate how the population exchange is ruled by internal decay rates and can be controlled by the intensity and detuning of a pump laser driving the red $\SI{688}{nm}$ transition between the levels $|5\rangle$ and $|6\rangle$. Maximal atom number in the green MOT is achieved with the $\SI{688}{\nano\meter}$ laser tuned to resonance and saturating the transition.

There is an asymmetry in the measured blue MOT fluorescence spectrum in Fig.~\ref{fig:Balancing}(b), which is probably due to the presence of the repump laser at $\SI{679}{\nano\meter}$. This laser, which is not accounted for in our simple model strongly saturates the transition causing saturation broadening, Autler-Townes splitting, and asymmetries if the laser is detuned from resonance.

\ParagSh{Impact of the external degrees of freedom} The external degrees of freedom are not included in the model presented above. Their impact on the system dynamics is accounted for in the extended theoretical framework presented in Appendix~\ref{sec:AppExternal}. There, a qualitative simulation of the dependency of the fluorescence on the magnetic field gradient and green laser detuning is provided. Against this background, we examine the double-peak structure emerging in the $\mathcal{F}_\text{blue}(\Delta_{34})$ data at given $B^\prime$ in Fig.~\ref{fig:RP_measurement}(a): the structure corresponds to the gRP configuration, where the atoms are cooled and trapped in two dimensions. As this configuration interpolates between the mere repump configuration and a two-dimensional MOT, it is likely at the origin of this structure, as shown in Fig.~\ref{fig:FluorescenceModel}.

\section{\label{sec:Conclusion}Conclusion and outlook}

In this paper we demonstrated that the atom number in a blue $^{88}\text{Sr}$ MOT can be enhanced by about one order of magnitude by aligning the beams of a $\SI{496}{\nano\meter}$ repump laser in a MOT configuration thus forming a continuously running two-color (blue-green) MOT. We showed that the system can be understood as an equilibrium between two subsystems represented by the steady states of the two MOTs. The point of equilibrium is then determined by internal decay processes, which can be manipulated by means of a $\SI{688}{\nm}$ pump laser, as we demonstrated experimentally and theoretically.

Our system represents an interesting platform for the realization of a continuous source of ultracold atoms paving the way towards uninterrupted operation of clocks and superradiant lasers. The method can also be applied to other repumping schemes and to other atomic species, such as Yb or Ca.

\section*{ACKNOWLEDGMENTS}

The authors thank Hidetoshi Katori for the joint initiation of the
Strontium project. We thank Christian Groß for valuable discussions, Brian Kasch for assistance in the lab, as well as Noriaki Ohmae and Florian Jessen for technical support. S.D. acknowledges funding from the Vector Stiftung via project number P2023-0105. Ph.W.C. acknowledges financial support from the Coordena\c{c}\~ao de Aperfeiçoamento de Pessoal de N\'ivel Superior (CAPES) (grant N$^\circ$\,23038.006739/2024-89) and the Funda\c{c}\~ao de Amparo \`a Pesquisa do Estado de S\~ao Paulo (FAPESP) (grant N$^\circ$\,2022/00209-6). J.F. and A.G. acknowledge financial support from the Deutsche Forschungsgemeinschaft through the Research Unit FOR 5413 (Grant No. 465199066).

\bigskip
\FloatBarrier
% \bibliography{Inputs/literature}

\begin{thebibliography}{10}

\bibitem{Nemitz16}
\href{https://doi.org/10.1038/nphoton.2016.20}{N.\,Nemitz, T.\,Ohkubo, M.\,Takamoto, I.\,Ushijima, M.\,Das, N.\,Ohmae, and H.\,Katori, \emph{Frequency ratio of {Y}b and {S}r clocks with $5\times 10^{-17}$ uncertainty at $150$ seconds averaging time}, Nature Phot. \textbf{10} (2016), 258.}

\bibitem{Norcia19}
\href{https://doi.org/10.1126/science.aay0644}{M.\,A. Norcia, A.\,W. Young, W.\,J. Eckner, E.\,Oelker, Jun Ye, and A.\,M. Kaufman, \emph{Seconds-scale coherence on an optical clock transition in a tweezer array}, Science \textbf{366} (2019), 93.}

\bibitem{Campbell17}
\href{https://doi.org/10.1126/science.aam5538}{S.\,L. Campbell, R.\,B. Hutson, G.\,E. Marti, A.\,Goban, N.\,Darkwah Oppong, R.\,L. McNally, L.\,Sonderhouse, J.\,M. Robinson, W.\,Zhang, B.\,J. Bloom, and J.\,Ye, \emph{A {F}ermi-degenerate three-dimensional optical lattice clock}, Science \textbf{358} (2017), 90.}

\bibitem{Ludlow15}
\href{https://doi.org/10.1103/RevModPhys.87.637}{A.\,D.\,Ludlow D., M.\,M. Boyd, Jun Ye, E.\,Peik, and P.\,O. Schmidt, \emph{Optical atomic clocks}, Rev. Mod. Phys. \textbf{87} (2015), 637.}

\bibitem{Takamoto05}
\href{https://doi.org/10.1038/nature03541}{M.\,Takamoto, Feng-Lei Hong, R.\,Higashi, and H.\,Katori, \emph{An optical lattice clock}, Nature \textbf{435} (2005), 321.}

\bibitem{Bennetts17}
\href{https://doi.org/10.1103/PhysRevLett.119.223202}{S.\,Bennetts, Chun-Chia Chen, B.\,Pasquiou, and F.\,Schreck, \emph{Steady-state magneto-optical trap with $100$-fold improved phase-space density}, Phys. Rev. Lett. \textbf{119} (2017), 223202.}

\bibitem{Dubey25}
\href{https://doi.org/10.1103/PhysRevResearch.7.013292}{S.\,Dubey, G.\,A. Kazakov, B.\,Heizenreder, Sheng Zhou, S.\,Bennetts, S.\,A. Sch\"affer, A.\,Sitaram, and F.\,Schreck, \emph{Modeling of a continuous superradiant laser on the sub-m{H}z $^1{S}_0\rightarrow\protect\,^3{P}_0$ transition in neutral {S}trontium-88}, Phys. Rev. Res. \textbf{7} (2025), 013292.}

\bibitem{Cao24}
\href{https://doi.org/10.1038/s41586-024-07913-z}{A.\,Cao, W.\,J. Eckner, Th.\,L. Yelin, A.\,W. Young, S.\,Jandura, Lingfeng Yan, Kyungtae Kim, G.\,Pupillo, Jun Ye, N.\,D. Oppong, and A.\,M. Kaufman, \emph{Multi-qubit gates and {S}chr\"odinger cat states in an optical clock}, Nature \textbf{634} (2024), 315.}

\bibitem{Stellmer10}
\href{https://doi.org/10.1103/PhysRevA.82.041602}{S.\,Stellmer, M.\,K. Tey, R.\,Grimm, and F.\,Schreck, \emph{{B}ose-{E}instein condensation of $^{86}${S}r}, Phys. Rev. A \textbf{82} (2010), 041602.}

\bibitem{Stellmer09}
\href{https://doi.org/10.1103/PhysRevLett.103.200401}{S.\,Stellmer, M.\,K. Tey, Bo\,Huang, R.\,Grimm, and F.\,Schreck, \emph{{B}ose-{E}instein condensation of {S}trontium}, Phys. Rev. Lett. \textbf{103} (2009), 200401.}

\bibitem{Bloch12}
\href{https://doi.org/10.1038/nphys2259}{I.\,Bloch, J.\,Dalibard, and S.\,Nascimb{\`e}ne, \emph{Quantum simulations with ultracold quantum gases}, Nature Phys. \textbf{8} (2012), 267.}

\bibitem{Bloch08}
\href{https://doi.org/10.1103/RevModPhys.80.885}{I.\,Bloch, J.\,Dalibard, and W.\,Zwerger, \emph{Many-body physics with ultracold gases}, Rev. Mod. Phys. \textbf{80} (2008), 885.}

\bibitem{Dunning16}
\href{https://doi.org/10.1088/0953-4075/49/11/112003}{F.\,B. Dunning, T.\,C. Killian, S.\,Yoshida, and J.\,Burgd\"orfer, \emph{Recent advances in {R}ydberg physics using alkaline-earth atoms}, J. Phys. B: At. Mol. Opt. Phys. \textbf{49} (2016), 112003.}

\bibitem{Gross17}
\href{https://doi.org/10.1126/science.aal3837}{C. Gross and I.\,Bloch, \emph{Quantum simulations with ultracold atoms in optical lattices}, Science \textbf{357} (2017), 995.}

\bibitem{Degen17}
\href{https://doi.org/10.1103/RevModPhys.89.035002}{C.\,L. Degen, F.\,Reinhard, and P.\,Cappellaro, \emph{Quantum sensing}, Rev. Mod. Phys. \textbf{89} (2017), 035002.}

\bibitem{McGuyer15}
\href{https://doi.org/10.1038/nphys3182}{B.\,H. McGuyer, M.\,McDonald, G.\,Z. Iwata, M.\,G. Tarallo, W.\,Skomorowski, R.\,Moszynski, and T.\,Zelevinsky, \emph{Precise study of asymptotic physics with subradiant ultracold molecules}, Nature Phys. \textbf{11} (2015), 32.}

\bibitem{Rivero23}
\href{https://doi.org/10.1088/1367-2630/acf954}{D.\,Rivero, C.\,A.\,Pessoa Jr., G.\,H. de\,Fran\c{c}a, R.\,C. Teixeira, S.\,Slama, and Ph.\,W. Courteille, \emph{Quantum resonant optical bistability with a narrow atomic transition: bistability phase diagram in the bad cavity regime}, New J. Phys. \textbf{25} (2023), 093053.}

\bibitem{XuX03}
\href{https://doi.org/10.1364/JOSAB.20.000968}{X.\,Xu, T.\,H. Loftus, J.\,L. Hall, A.\,Gallagher, and J.\,Ye, \emph{Cooling and trapping of atomic {S}trontium}, J. Opt. Soc. Am. B \textbf{20} (2003), 968.}

\bibitem{Drozdowski97}
\href{https://doi.org/10.1007/s004600050300}{R.\,Drozdowski, M.\,Ignaciuk, J.\,Kwela, and J.\,Heldt, \emph{Radiative lifetimes of the lowest $^3{P}_1$ metastable states of {C}a and {S}r}, Z. Phys. D: Atoms Mol. Clusters \textbf{41} (1997), 125.}

\bibitem{Stellmer14b}
\href{https://doi.org/10.1142/9789814590174_0001}{S.\,Stellmer, F.\,Schreck, and T.\,C. Killian, \emph{Degenerate quantum gases of {S}trontium}, Annu. Rev. Cold At. Mol., May 2014, 1-80.}

\bibitem{Yasuda04}
\href{https://doi.org/10.1103/PhysRevLett.92.153004}{M.\,Yasuda and H.\,Katori, \emph{Lifetime measurement of the $^3{P}_2$ metastable state of {S}trontium atoms}, Phys. Rev. Lett. \textbf{92} (2004), 153004.}

\bibitem{Dineen99}
\href{https://doi.org/10.1103/PhysRevA.59.1216}{T.\,P. Dinneen, K.\,R. Vogel, E.\,Arimondo, J.\,L. Hall, and A.\,Gallagher, \emph{Cold collisions of {S}r$^*-${S}r in a magneto-optical trap}, Phys. Rev. A \textbf{59} (1999), 1216.}

\bibitem{Poli06b}
\href{https://doi.org/10.1016/j.saa.2005.10.024}{N. Poli, G. Ferrari, M. Prevedelli, F. Sorrentino,  R.\,E. Drullinger, G.M. Tino, \emph{Laser sources for precision spectroscopy on atomic {S}trontium}, Spectrochimica Acta Part A: Mol. Biomol. Spectroscopy \textbf{63} (2006), 981.}

\bibitem{Bowden19}
\href{https://doi.org/10.1038/s41598-019-48168-3}{W.\,Bowden, R.\,Hobson, I.\,R. Hill, A.\,Vianello, M.\,Schioppo, A.\,Silva, H.\,S. Margolis, P.\,E.\,G. Baird, and P.\,Gill, \emph{A pyramid {MOT} with integrated optical cavities as a cold atom platform for an optical lattice clock}, Sci. Rep. \textbf{9} (2019), 11704.}

\bibitem{Camargo-17}
\href{https://doi.org/}{F.\,Camargo, \emph{Rydberg molecules and polarons in ultracold {S}trontium gases}, Dissertation, Rice University, Houston, USA, 2017.}

\bibitem{HuF19}
\href{https://doi.org/10.1103/PhysRevA.99.033422}{F.\,Hu, I.\,Nosske, L.\,Couturier, C.\,Tan, C.\,Qiao, P.\,Chen, Y.\,H. Jiang, B.\,Zhu, and M.\,Weidem\"uller, \emph{Analyzing a single-laser repumping scheme for efficient loading of a {S}trontium magneto-optical trap}, Phys. Rev. A \textbf{99} (2019), 033422.}

\bibitem{Kurosu92}
\href{https://doi.org/10.1143/JJAP.31.908}{T.\,Kurosu and F.\,Shimizu, \emph{Laser cooling and trapping of alkaline earth atoms}, Jap. J. Appl. Phys. \textbf{31} (1992), 908.}

\bibitem{Moriya18}
\href{https://doi.org/10.1088/2399-6528/aaf662}{P.\,H. Moriya, M.\,O. Ara\'ujo, F.\,Tod\~a o, M.\,Hemmerling, H.\,Ke{\ss}ler, R.\,F. Shiozaki, R.\,Celistrino Teixeira, and Ph.\,W. Courteille, \emph{Comparison between $403$nm and $497$nm repumping schemes for {S}trontium magneto-optical traps}, J. of Phys. Commun. \textbf{2} (2018), 125008.}

\bibitem{Mickelson09}
\href{https://doi.org/10.1088/0953-4075/42/23/235001}{P.\,G. Mickelson, Y.\,N. de\,Escobar\,Martinez, P.\,Anzel, B.\,J. DeSalvo, S.\,B. Nagel, A.\,J. Traverso, M.\,Yan, and T.\,C. Killian, \emph{Repumping and spectroscopy of laser-cooled sr atoms using the $(5s5p)\,^3{P}_2–(5s4d)\,^3{D}_2$ transition}, J. Phys. B: At., Mol. and Opt. Phys. \textbf{42} (2009), 235001.}

\bibitem{Mills17}
\href{https://doi.org/10.1103/PhysRevA.96.033402}{M.\,Mills, P.\,Puri, Y.\,Yu, A.\,Derevianko, Ch. Schneider, and E.\,R. Hudson, \emph{Efficient repumping of a {C}a magneto-optical trap}, Phys. Rev. A \textbf{96} (2017), 033402.}

\bibitem{Poli05}
\href{https://doi.org/10.1103/PhysRevA.71.061403}{N.\,Poli, R.\,E. Drullinger, G.\,Ferrari, J.\,L\'eonard, F.\,Sorrentino, and G.\,M. Tino, \emph{Cooling and trapping of ultracold {S}trontium isotopic mixtures}, Phys. Rev. A \textbf{71} (2005), 061403.}

\bibitem{Poli06}
\href{https://doi.org/10.1016/j.saa.2005.10.024}{N.\,Poli, G.\,Ferrari, M.\,Prevedelli, F.\,Sorrentino, R.\,E. Drullinger, and G.\,M. Tino, \emph{Laser sources for precision spectroscopy on atomic {S}trontium}, Spectrochim. Acta A: Mol. Biomol. Spectrosc. \textbf{63} (2006), 981.}

\bibitem{Samland24}
\href{https://doi.org/10.1103/PhysRevRes.6.013319}{J.\,Samland, S.\,Bennetts, Chun-Chia Chen, R.\,G. Escudero, F.\,Schreck, and B.\,Pasquiou, \emph{Optical pumping of $5s4d\,^1{D}_2$ {S}trontium atoms for laser cooling and imaging}, Phys. Rev. Res. \textbf{6} (2024), 013319.}

\bibitem{Schkolnik2019}
\href{https://doi.org/10.1088/1555-6611/aaffc8}{V.\,Schkolnik, O.\,Fartmann, and M.\,Krutzik, \emph{An extended-cavity diode laser at $497$nm for laser cooling and trapping of neutral {S}trontium}, Laser Phys. \textbf{29} (2019), 035802.}

\bibitem{Sorrentino06}
\href{https://doi.org/10.1142/S0217984906011682}{F.\,Sorrentino, G.\,Ferrari, N.\,Poli, R.\,Drullinger, and G.\,M. Tino, \emph{Laser cooling and trapping of atomic {S}trontium for ultracold atom physics, high-precision spectroscopy and quantum sensors}, Mod. Phys. Lett. B \textbf{20} (2006), 1287.}

\bibitem{Stellmer14}
\href{https://doi.org/10.1103/PhysRevA.90.022512}{S.\,Stellmer and F.\,Schreck, \emph{Reservoir spectroscopy of $5s5p\,^3{P}_2–5s\,nd\, ^3{D}_{1,2,3}$ transitions in {S}trontium}, Phys. Rev. A \textbf{90} (2014), 022512.}

\bibitem{Takamoto2020}
\href{https://doi.org/10.1038/s41566-020-0619-8}{M.\,Takamoto, I.\,Ushijima, N.\,Ohmae, T.\,Yahagi, K.\,Kokado, H.\,Shinkai, and H.\,Katori, \emph{Test of general relativity by a pair of transportable optical lattice clocks}, Nature Phot. \textbf{14} (2020), 411.}

\bibitem{ZhangS20}
\href{https://doi.org/10.48550/arXiv.2007.10465}{S.\,Zhang, P.\,Ramchurn, M.\,Menchetti, Q.\,Ubaid, J.\,Jones, K.\,Bongs, and Y.\,Singh, \emph{Novel repumping on $^3{P}_0\rightarrow{^3}{D}_1$ for {S}r magneto-optical trap and {L}and\'e g factor measurement of $^3{D}_1$}, J. Phys. B: At., Mol., and Opt. Phys. \textbf{53} (2020), 235301.}

\bibitem{Barker15}
\href{https://doi.org/10.1103/PhysRevA.92.043418}{D.\,S. Barker, B.\,J. Reschovsky, N.\,C. Pisenti, and G.\,K. Campbell, \emph{Enhanced magnetic trap loading for atomic {S}trontium}, Phys. Rev. A \textbf{92} (2015), 043418.}

\bibitem{Akatsuka21}
\href{https://doi.org/10.1103/PhysRevA.103.023331}{T.\,Akatsuka, K.\,Hashiguchi, T.\,Takahashi, N.\,Ohmae, M.\,Takamoto, and H.\,Katori, \emph{Three-stage laser cooling of {S}r atoms using the $5s5p\,^3{P}_2$ metastable state below doppler temperatures}, Phys. Rev. A \textbf{103} (2021), 023331.}

\bibitem{Negyedi25}
\href{https://arxiv.org/pdf/2507.02693}{M.\,J. Negyedi, S.\,Deutschle, F.\,Jessen, J.\,Fort\'agh, and L.\,S\'ark\'any, \emph{Sub-{D}oppler cooling of bosonic {S}trontium in a two-color {MOT}}, arXiv:2507.02693 (2025).}

\bibitem{Note1}
For simplicity we ignored the intermediate $5s4d\protect \,^1D_2$ level in the blue-to-green MOT decay path. However, the decay rate $5s5p\protect\,^1P_1\rightarrow 5s4d\protect \,^1D_2$ is faster than $5s4d\protect\,^1D_2\rightarrow 5s5p\protect \,^3P_{1,2}$ so that a non-negligible number of atoms can be shelved in that intermediate level.

\bibitem{Kawasaki16}
\href{https://doi.org/10.1088/0953-4075/48/15/155302}{A.\,Kawasaki, B.\,Braverman, Q.\,Q. Yu, and V.\,Vuletic, \emph{Two-color magneto-optical trap with small magnetic field for {Y}tterbium}, J. Phys. B: At. Mol. Opt. Phys. \textbf{48} (2015), 155302.}

\bibitem{Hoschele23}
\href{https://doi.org/10.1103/PhysRevApplied.19.064011}{J.\,H\"oschele, S.\,Buob, A.\,Rubio-Abadal, V.\,Makhalov, and L.\,Tarruell, \emph{Atom-number enhancement by shielding atoms from losses in {S}trontium magneto-optical traps}, Phys. Rev. Appl. \textbf{19} (2023), 064011.}

\bibitem{Lett89}
\href{https://doi.org/10.1088/0031-8949/1991/T34/003}{P.\,D. Lett, W.\,D. Phillips, S.\,L. Rolston, C.\,E. Tanner, R.\,N. Watts, and C.\,I. Westbrook, \emph{Optical molasses}, J. Opt. Soc. Am. B \textbf{6} (1989), 2084.}

\bibitem{Homa24}
\href{https://doi.org/10.1515/zna-2024-0144}{Tin-Lun Ho and Sungkit Yip, \emph{Choi representation of completely positive maps in brief}, Z. Naturforsch. \textbf{79} (2024), 1123.}


\end{thebibliography}
\bibliographystyle{amsplain}

\FloatBarrier
\appendix

\section{Appendix}

\subsection{\label{sec:AppDecays}Decay rates between the MOTs}

The decay rate from $^1P_1$ via $^1D_2$ to $^3P_1$ is,
\begin{small}\begin{align}\label{eq:App01}
    \Gamma_{25} & = \Gamma\footnotesize{\Mtz{5s5p\,^1P_1 \\ \rightarrow 5s5p\,^3P_1}}\\
    & = \Gamma\footnotesize{\Mtz{5s5p\,^1P_1 \\ \rightarrow 5s4d\,^1D_2}}\parallel
        \Gamma\footnotesize{\Mtz{5s4d\,^1D_2 \\ \rightarrow 5s5p\,^3P_1}}
    = (2\pi)\,\SI{159}{\Hz}\nonumber \, .
\end{align}\end{small}
The symbol $\parallel$ denotes summation of the reciprocal values, $\Gamma_a \parallel \Gamma_b \equiv (\Gamma_a^{-1}+\Gamma_b^{-1})^{-1}$. The individual decay rates are taken from Fig.~\ref{fig:Levelscheme}. 
The decay rate from $^1P_1$ via $^1D_2$ to $^3P_2$ is,
\begin{small}\begin{align}\label{eq:App02}
    \Gamma_{23} & = \Gamma\footnotesize{\Mtz{5s5p\,^1P_1 \\ \rightarrow 5s5p\,^3P_2}}\\
    & = \Gamma\footnotesize{\Mtz{5s5p\,^1P_1 \\\rightarrow 5s4d\,^1D_2}}\parallel
        \Gamma\footnotesize{\Mtz{5s4d\,^1D_2 \\ \rightarrow 5s5p\,^3P_2}}
    = (2\pi)\,\SI{90}{\Hz}\nonumber \, .
\end{align}\end{small}
The decay rate from $^3D_3$ to $^3P_1$ is,
\begin{small}\begin{align}\label{eq:App03}
    \Gamma_{45} & = \Gamma\footnotesize{\Mtz{5s5d\,^3D_3 \\\rightarrow 5s5p\,^3P_1}}\\
    & = \Gamma\footnotesize{\Mtz{5s5d\,^3D_3 \\ \rightarrow 5s6p\,^3P_2}}\parallel \footnotesize{\Mtz{
        ~~\Gamma(\rightarrow 5s4d\,^1D_2)\parallel\Gamma(\rightarrow 5s5p\,^3P_1) \\
        +\Gamma(\rightarrow 5s6s\,^3S_1)\parallel\Gamma(\rightarrow 5s5p\,^3P_1) \\
        +\Gamma(\rightarrow 5s4d\,^3D_2)\parallel\Gamma(\rightarrow 5s5p\,^3P_1) \\
        +\Gamma(\rightarrow 5s4d\,^3D_1)\parallel\Gamma(\rightarrow 5s5p\,^3P_1)}}\nonumber\\
    & = (2\pi)\,\SI{26.3}{\kilo\Hz}\nonumber \, .
\end{align}\end{small}
Finally, the decay rate from $^3P_1$ to $^1S_0$ is,
\begin{equation}\label{eq:App04}
    \Gamma_{15} = \Gamma\footnotesize{\Mtz{5s5p\,^3P_1 \\ \rightarrow 5s^2\,^1S_0}}
    = (2\pi)\,\SI{7.4}{\kHz}\,.
\end{equation}

\subsection{\label{sec:alternativeRepumps}Overview of repumping schemes for the blue Strontium MOT}

The following list provides a more detailed overview of the alternative repump schemes that have been reported in the literature:
\begin{itemize}[itemsep=0pt, parsep=0pt, topsep=0pt, partopsep=0pt]
    \item $5s5p\,^3P_2\rightarrow 5s5d\,^3D_2$ at $\SI{497}{\nano\meter}$~\cite{Bowden19,Stellmer14,Poli05,Sorrentino06,Stellmer09,Stellmer14b,Poli06,Schkolnik2019, Poli06b,Moriya18}
    \item $5s5p\,^3P_2 \rightarrow 5p^2\,^3P_2$ at $\SI{481}{\nano\meter}$~\cite{HuF19, Camargo-17}
    \item $5s5p\,^3P_2 \rightarrow 5s6d\,^3D_2$ at $\SI{403}{\nano\meter}$~\cite{Stellmer14,Moriya18}
    \item $5s5p\,^3P_2 \rightarrow 5s4d\,^3D_2$ at $\SI{3.012}{\micro\meter}$~\cite{Mickelson09}
    \item $5s5p\,^3P_0 \rightarrow 5s4d\,^3D_1$ at $\SI{2.6}{\micro\meter}$~\cite{ZhangS20}~\item $5s4d\,^1D_2 \rightarrow 5s8p\,^1P_1$ at $\SI{448}{\nano\meter}$~\cite{Samland24,Mills17}
    \item $5s4d\,^1D_2 \rightarrow 5s6p\,^1P_1$ at $\SI{717}{\nano\meter}$~\cite{Kurosu92} 
\end{itemize}

With the excited states of those transitions which have a negligible decay probability into the metastable $5s5p\,^3P_0$ state, no additional laser is required for its depletion.

\subsection{\label{sec:AppExternal}Model for the external degrees of freedom}

In contrast to the constants and experimental parameters determining the steady state populations of all subsystems, which are known or easily measured, the loading and loss rates also depend in a complicated way on experimental settings or adjustments. In order to interpret our measurements in Fig.~\ref{fig:RP_measurement}, we thus have to come up with plausible models on how $R_\text{load}$ and $\Gamma_\text{color}$ depend on specific experimental parameters. 
 
In principle, simulations of atomic trajectories exposed to cooling laser beams and magnetic fields can help determining optimal laser detunings and magnetic field gradients~\cite{Kawasaki16}. Many effect come into play: (i)~The atomic cloud is subject to severe inhomogeneous broadenings due to its velocity and spatial distribution over the magnetic field region, which is not captured by the Bloch-rate equations model. (ii)~The cooling dynamics (in particular sub-Doppler cooling on the green transition) presupposes a multilevel structure and acts back on the internal state dynamics. (iii)~Radiation pressure limits the attainable density. (iv)~It is known that MOT loading can be enhanced if a fraction of the atoms is shelved in dark states~\cite{Hoschele23}.

Because of these complications, trajectory simulations are, however, rarely quantitatively satisfactory, mostly intransparent, and beyond the scope of this paper. Instead, we propose in the following a simplistic model heuristically assuming certain dependencies of $R_\text{load}$ and $\Gamma_\text{color}$ on specific experimental parameters. These dependencies then resume the entire MOT cooling dynamics.

\ParagSh{Forces in a MOT} We start by investigating the dynamics of the blue and green MOT separately in terms of external degrees of freedom. In one dimension the force on an atom in the MOT results from two counterpropagating laser beams $F_\pm$. It can be described as a function of the velocity $v$ along the axis of the laser propagation $z$ as~\cite{Lett89}:
\begin{equation}\label{eq:Force.MOT}
    F_\pm(v,z) = \pm\frac{\hbar k_{ij}\Gamma_{ij}}{2}\frac{s_{ij}\,\theta(z+w_{ij})
    \theta(w_{ij}-z)}{1+s_{ij}+4\left(\frac{\Delta_{ij}\pm k_{ij}v\mp\mu B'z/\hbar}{\Gamma_{ij}}\right)^2}~,
\end{equation}
with $(ij)=(12)$ for the blue MOT and $(ij)=(34)$ for the green MOT. $\Gamma_{ij}$ is the linewidth of the MOT-transition, $\Delta_{ij}$ is the light detuning to the resonance, $k_{ij}=2\pi/\lambda_{ij}$ is the wavenumber with the wavelength $\lambda_{ij}$, $w_{ij}$ is the beam radius, $s_{ij}$ is the saturation parameter, and $B'$ is the gradient of the magnetic field. Note that the Landé factors are $g_J(^1S_0)=1=g_{J'}(^1P_1)$ for the blue MOT levels and $g_J(^3P_2)=\tfrac{3}{2}$ and $g_{J'}(^3D_3)=\tfrac{4}{3}$ for the green MOT levels, so that transitions between fully stretched Zeeman substates ($|m_J|=J$) have identical differential magnetic moments,
\begin{equation}
	\mu = \mu_\text{B}(g_{J'}m_{J'}-g_Jm_J) = \mu_\text{B}~.
\end{equation}
The Heaviside step function $\theta$ accounts for the fact that, although along the $z$-axis the MOT force is present at all distances, atoms outside the radius of the transverse MOT beams $|z|>w_{ij}$ drift out sideways of the MOT region.
\begin{figure}[htbp]
    \centering
    \includegraphics[width=8.7cm]{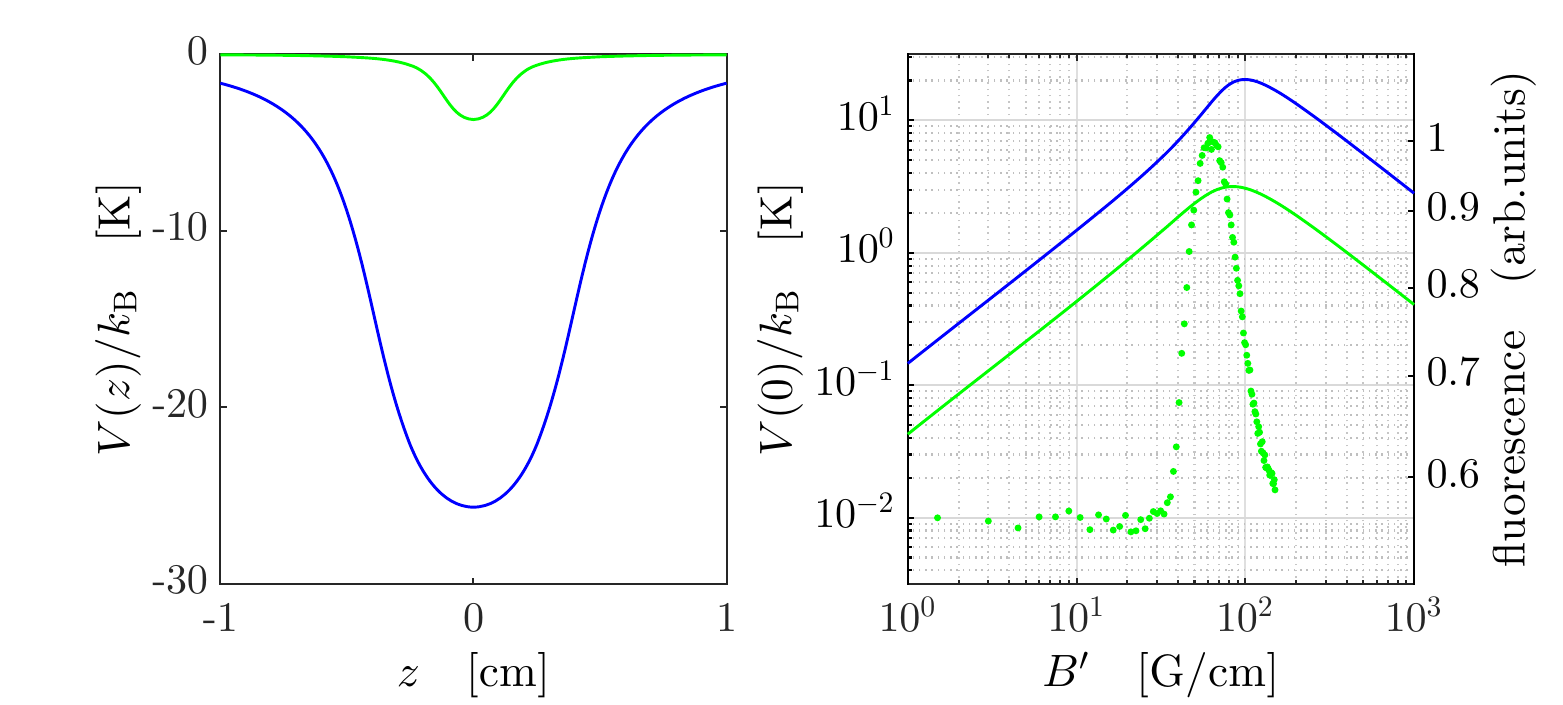}
    \caption{\small{(a)~Trapping potential for the blue and green MOT for the parameters: $s_{12}=2$ and $s_{34}=6.6$, $1/e^2$-radius $w_{12}=\SI{6}{\milli\meter}$ and  $w_{34}=\SI{2.3}{\milli\meter}$, $\Delta_{12}=-2\, \Gamma_{12}$ and $\Delta_{34}=-2\,\Gamma_{34}$.
    (b)~Depth of the trapping potential according to Eq.~\eqref{eq:App21} for both MOTs as a function of the magnetic field gradient. The symbols denote measurements of the green MOT fluorescence as a function of the magnetic field gradient.}}
    \label{fig:Fig_MOTPotential}
\end{figure}

\ParagSh{Trap depth} We calculate the trapping potential for already cooled and trapped atoms ($v=0$) by integrating Eq.~(\ref{eq:Force.MOT}) along $z$. The spatial limitation of the MOT is given here by the cooling beams intersecting the trap in 3D. Therefore, we integrate Eq.~(\ref{eq:Force.MOT}) over the beam radius along $z$ and obtain,
\begin{small}\begin{align}\label{eq:Force.Pot}
    V(z) & = \int_{-\infty}^z[F_+(0,z')+F_-(0,z')]dz'\\
    & = \frac{\hbar^2k_{ij}}{4}\frac{\Gamma_{ij}^2s_{ij}}{\mu B'\sqrt{1+s_{ij}}}
    \sum_\pm\left(\arctan\frac{2\left(\Delta_{ij}\pm w_{ij}\frac{\mu}{\hbar}B'\right)}{\Gamma_{ij}\sqrt{1+s_{ij}}}\right.\nonumber\\
    & \left. \hspace{2cm} -\arctan \frac{2\left(\Delta_{ij}\pm\min(w_{ij},z)\frac{\mu}{\hbar}B'\right)}{\Gamma _{ij}\sqrt{1+s_{ij}}}\right)~.\nonumber
\end{align}\end{small}
Evaluation of this expression at $z=0$ yields the potential depth,
\begin{align}\label{eq:App21}
    V(0) & = \frac{\hbar^2k_{ij}}{4}\frac{\Gamma_{ij}^2s_{ij}}{\mu B'\sqrt{1+s_{ij}}}\times\\
    & \times \sum_{\xi=-1,0,0,1}(-1)^\xi\arctan\frac{\Delta_{ij}+\xi w_{ij}\tfrac{\mu}{\hbar}B'}{\tfrac{\Gamma_{ij}}{2}\sqrt{1+s_{ij}^2}}~.\nonumber
\end{align}
The trapping potentials and their respective depths for the blue and green MOTs are shown in Fig.~\ref{fig:Fig_MOTPotential}.

Measurements of the green-MOT fluorescence as a function of the magnetic-field gradient [see Fig.~\ref{fig:Fig_MOTPotential}(b)] show that the MOT fluorescence reaches its maximum at a gradient close to the theoretically predicted value for the maximum trap depth.

\begin{figure}[htbp]
    \centering
    \includegraphics[width=7cm]{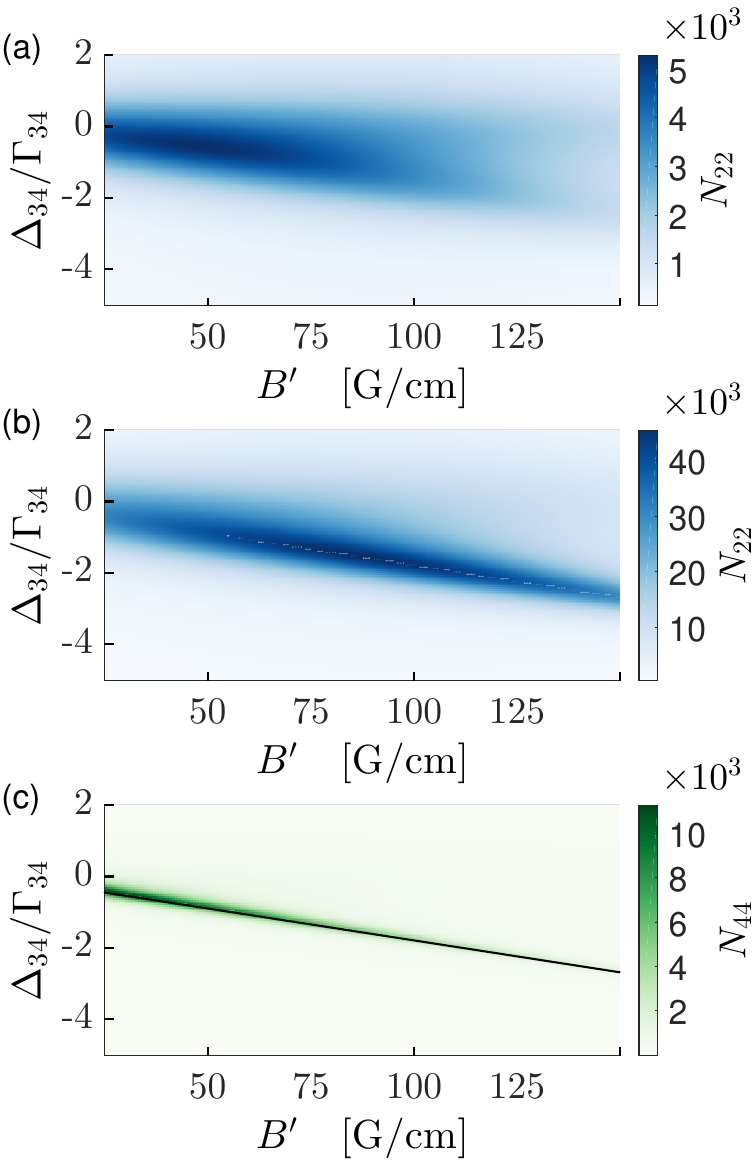} 
    \caption{\small{Simulated blue and green fluorescences obtained in the gRP and the gMOT configurations as a function of magnetic field gradient and green laser detuning. (a)~Blue fluorescence in gRP configuration, (b)~blue fluorescence in gMOT configuration, and (c)~green fluorescence in gMOT configuration. The black solid line is calculated from Eq.~\eqref{eq:App23}.}}
    \label{fig:FluorescenceModel}
\end{figure}

\ParagSh{Optimum detuning} We concentrate on the values of the gradients at which the potentials are deepest dependent on the beam radius, detuning, and intensity. At these gradients, the MOTs are expected to operate nearly optimal. Comparing the maximum gradients for the blue and green MOT, we note that for our parameters they nearly coincide. To determine for a given gradient $B'$ the detuning $\Delta _{ij}^{(m)}$ at which the potential is deepest, we solve the equation $0=dV(0)/d\Delta_{ij}$ and obtain
\begin{align}\label{eq:App23}
    \Delta_{ij}^{(m)} = -\sqrt{\frac{(w_{ij}\mu B')^2}{3\hbar^2}+\frac{(1+s_{ij})\Gamma_{ij}^2}{12}}\\
    \overset{B'\rightarrow\infty}{\longrightarrow} -\sqrt{\frac{1}{3}}\frac{w_{ij}\mu B'}{\hbar}~.\nonumber
\end{align}
The dependence of $\Delta_{ij}^{(m)}$ on $B'$ is shown as solid black line in Figs.~\ref{fig:FluorescenceModel}(c) and~\ref{fig:RP_measurement}(d).

\ParagSh{Interpretation} The MOT relies on the Zeeman shift to provide a spatially dependent restoring force. If the gradient is too weak, the restoring force is small. Then atoms with modest transverse velocity escape before being slowed. Hence, the capture volume shrinks entailing lower efficiency. If the gradient is too strong, the Zeeman shift moves atoms quickly out of resonance as they move away from the center. That is, only a very narrow spatial region remains resonant, the effective capture velocity decreases again leading to lower efficiency.

Thus, there is an optimal gradient that can roughly be estimated via the Zeeman shift across the MOT beam radius (setting $z=w$ in the above formula). 

\ParagSh{Modelling the loading rate} As a very rough model, we assume that the loading rate follows a Gaussian profile centered at the optimal magnetic-field gradient,
\begin{align}\label{eq:App24}
	R_\text{load} \sim R_\text{load,0}\,e^{-(B'-B'_\text{m,blue})^2/(\Delta B'_{\text{blue}})^2}
\end{align}
where
\begin{equation}\label{eq:App25}
    B'_\text{m,blue} = -\sqrt{3}\frac{\hbar\Delta_{12}}{\mu w_{12}}
\end{equation}
and the width of the Gaussian distribution is estimated to roughly $\Delta B'_{\text{blue}}\approx\SI{100}{G/cm}$ from the measurements exhibited in Figs.~\ref{fig:RP_measurement}.

\ParagSh{Modelling the loss rates} The experimental observations shown in Fig.~\ref{fig:RP_measurement} indicate that the present model does not capture all relevant physical mechanisms. In particular, it does not include dynamical effects associated with the MOT. The observed asymmetry around the atomic resonance with respect to the green-laser detuning may therefore originate from MOT dynamics and thus involve the external degrees of freedom. This interpretation is supported by the distinctly different behavior observed for the gRP and gMOT configurations under otherwise identical conditions. The missing ingredient is therefore likely related to the impact of the green light on the external degrees of freedom.

The only handle offered by our simplistic model is a functional dependence of the loss rate $\Gamma_\text{gr:rd}=\Gamma_\text{gr:rd}(\Delta_{34},B')$, which is shaped by the green laser configuration. Physically, this is motivated by the observation that the green MOT (gMOT configuration) is only capable of confining atoms at certain detunings and gradients. On the other hand, the two-dimensional MOT (gRP configuration) has a much lower confinement capability, the atoms are partially free. We therefore make the following ansatz for the green loss rate,
\begin{align}\label{eq:Gamma.RP}
    \Gamma_\text{gr:rd}(\Delta_{34},B') & = \Gamma_\text{free}\\
    & + \Gamma_\text{trap}(1-e^{-(B'-B'_\text{m,green})^2/(\Delta B'_\text{gr:rd})^2})\nonumber
\end{align}
where
\begin{equation}\label{eq:App25}
    B'_\text{m,green} = -\sqrt{3}\frac{\hbar\Delta_{34}}{\mu w_{34}}~.
\end{equation}

\ParagSh{Simulations} The panels in Fig.~\ref{fig:FluorescenceModel} show the excited state populations $\rho_{22}\propto\mathcal{F}_\mathrm{blue}$ and $\rho_{44}\propto\mathcal{F}_\mathrm{green}$ as a function of the green laser detuning $\Delta_{34}$ and the magnetic field gradient $B'$. The green loss rates have been calculated from Eq.~\eqref{eq:Gamma.RP} with $\Delta B'_\text{green}$ estimated to roughly $\SI{100}{G/cm}$ from the measurements exhibited in Figs.~\ref{fig:RP_measurement}. Also, we assume $\Gamma_\text{trap}^\text{(gMOT)}=200\,\text{s}^{-1}$ and $\Gamma_\text{free}^\text{(gMOT)}=1\,\text{s}^{-1}$,  $\Gamma_\text{trap}^\text{(gRP)}=1000\,\text{s}^{-1}$ and $\Gamma_\text{free}^\text{(gRP)}=300\,\text{s}^{-1}$, respectively.

The simulations qualitatively reproduce the observations shown in Fig.~\ref{fig:RP_measurement}: (1)~Increasing $\Gamma_\text{gr:rd}$ (e.g.~switching from the gRP to the gMOT configuration) dramatically reduces the blue fluorescence. Our simulations confirm the experimentally observed gain of $10$ in atom numbers [see Fig.~\ref{fig:RP_measurement}(c)]. (2)~The overall fluorescence landscape is similar in both, measurement and simulation, including gradients and double-peak structures.

\subsection{\label{sec:AppSatint}Saturation intensities and parameters}

The saturation intensity between individual Zeeman sublevels with Clebsch-Gordon coefficient $\text{CGC}=1$ is given by,
\begin{equation}\label{eq:App11}
    s = \frac{I}{I_\text{sat}} ~~~\text{where}~~~ 
    I = \frac{2P}{\pi w_0^2} ~~~\text{and}~~~ 
    I_\text{sat} = \frac{2\pi^2c\hbar}{3\lambda^3}\Gamma~,
\end{equation}
where $\Gamma$ and $\lambda$ are the linewidth and wavelength of the considered transition. We have $I_{\text{sat,461}}=\SI{38}{\milli\W/cm^2}$, $I_\text{sat,496}=\SI{9.1}{\milli\W/cm^2}$, and $I_\text{sat,689}=\SI{2.8}{\micro\W/cm}^2$. The irradiated saturation parameters in the experiment are listed below:\\\\
\begin{tabular}[c]{l|c|c|c|c}
    \textbf{measurement} & \textbf{transition} & \textbf{waist} & \textbf{total power} & \textbf{$s$}\\\hline
    Fig.~\ref{fig:RP_measurement} & $\SI{461}{nm}$ & $\SI{2.9}{\milli\meter}$ & $\SI{40}{\milli\watt}$ & $8.0$\\
    Fig.~\ref{fig:RP_measurement}(a-b) & $\SI{496}{nm}$ & $\SI{2.2}{\milli\meter}$ & $\SI{2.8}{\milli\watt}$ & $4.0$\\
    Fig.~\ref{fig:RP_measurement}(d) & $\SI{496}{nm}$ & $\SI{2.2}{\milli\meter}$ & $\SI{8.5}{\milli\watt}$ & $12.3$\\
    Fig.~\ref{fig:Balancing} & $\SI{461}{nm}$ & $\SI{6.0}{\milli\meter}$ & $\SI{44}{\milli\watt}$ & $2.0$\\
    Fig.~\ref{fig:Balancing} & $\SI{496}{nm}$ & $\SI{2.3}{\milli\meter}$ & $\SI{5}{\milli\watt}$ & $6.6$\\
    Fig.~\ref{fig:Balancing} & $\SI{688}{nm}$ & $\SI{1.8}{\milli\meter}$ & $\SI{2.6}{\milli\watt}$ & $31$
\end{tabular}

The beam waist of the $\SI{688}{\nano\meter}$ laser is $\SI{.9}{\milli\meter}$. The beam waist of the $\SI{679}{\nano\meter}$ laser is about $\SI{5}{\milli\meter}$.

\subsection{\label{sec:AppDerive}Derivation of the open Bloch equations}

We describe the system dynamics via the standard master equation,
\begin{equation}\dot{\hat N} = \tfrac{\imath}{\hbar}[\hat N,\hat H]+\tfrac{1}{2}\sum_{i,j}\Gamma_{ij}\mathcal{L}_{\hat\sigma_{ij}}\hat N~.
\end{equation}
The density matrix normalized to the total atom number is,
\begin{small}\begin{equation}\label{eq:The01}
    \hat N = \left(\begin{array}[c]{cccccccccccccc}
        N_{11} & N_{12} & 0 & 0 & 0 & 0 \\
        N_{21} & N_{22} & 0 & 0 & 0 & 0 \\
        0 & 0 & N_{33} & N_{34} & 0 & 0 \\
        0 & 0 & N_{43} & N_{44} & 0 & 0 \\
        0 & 0 & 0 & 0 & N_{55} & N_{56} \\
        0 & 0 & 0 & 0 & N_{65} & N_{66} \\
    \end{array}\right) \, .
\end{equation}\end{small}
The coherent part of the evolution can be handled by a Hamiltonian
\begin{small}\begin{equation}\label{eq:The02}
    \hat H = \Mtz{\hat H_{|1\,2\rangle} & 0 & 0 \\
        0 & \hat H_{|3\,4\rangle} & 0 \\
        0 & 0 & \hat H_{|5\,6\rangle}}~,
\end{equation}\end{small}
which is composed of three orthogonal subspace Hamiltonians
\begin{equation}\label{eq:The03}
    \hat H_{|i\,j\rangle} = \Mtz{0 & \tfrac{1}{2}\Omega_{ij} \\
    \tfrac{1}{2}\Omega_{ij} & -\Delta_{ij}}~.
\end{equation}

The incoherent coupling through spontaneous decay processes from states $|j\rangle$ to $|i\rangle$ occurring at rates $\Gamma_{ij}$ is handled by deexcitation operators $\hat\sigma_{ij}$. Choi-Jamio\l skowski vectorization of all operators then transforms the master equation into the open Bloch equations~\eqref{eq:mod02} presented in the main text \cite{Homa24}.

\vspace{5cm}

\end{document}